\documentclass[12pt,a4paper]{article}
\usepackage{graphicx}
\usepackage{amssymb}
\usepackage{amsmath}

\begin{document}

\title{Theory of neutrino-atom collisions: the history, present
status and BSM physics}

\author{Konstantin A. Kouzakov\\
{\small \it Department of Nuclear Physics and Quantum Theory of Collisions,}\\
{\small \it Faculty of
Physics, Lomonosov Moscow State University,}\\ {\small \it 119991 Moscow, Russia}\\
\texttt{\small kouzakov@srd.sinp.msu.ru}\\
\and
Alexander I. Studenikin$^{1, 2}$\\
{\small \it $^1$ Department of Theoretical Physics,}\\ {\small \it Faculty
of Physics, Lomonosov
Moscow State University,}\\ {\small \it 119991 Moscow, Russia}\\
{\small \it $^2$ Joint Institute for Nuclear Research, 141980 Dubna, Russia}\\
\texttt{\small studenik@srd.sinp.msu.ru}}


\maketitle

\begin{abstract}
An overview of the current theoretical studies on neutrino-atom scattering
processes is presented. The ionization channel of these processes, which
is studied in experiments searching for neutrino magnetic moments, is
brought into focus. Recent developments in the theory of atomic ionization
by impact of reactor antineutrinos are discussed. It is shown that the
stepping approximation is well applicable for the data analysis
practically down to the ionization threshold.
\end{abstract}
\section{\label{intro}Introduction}
In particle physics, the neutrino plays a remarkable role of a ``tiny''
particle. The scale of neutrino mass $m_\nu$ is much lower than that of
the charged fermions ($m_{\nu_f}\ll m_f$, $f = e, \mu, \tau$). Interaction
of neutrinos with matter is extremely weak as compared to that in the case
of other known elementary fermions. According to the Standard Model (SM),
it can be mediated only via exchange of the $W^\pm$ and $Z^0$ bosons.
However, the recent development of our knowledge of neutrino mixing and
oscillations, supported by the discovery of flavor conversions of
neutrinos from different sources
(see~\cite{giunti_book,gonzalez-garcia08,bilenky11,xz_book}),
substantiates the assumption that neutrinos can possess electromagnetic
properties and, hence, take part in electromagentic interactions (see, for
instance, the review articles~\cite{giunti09,broggini12,Giunti:2014ixa}). These
properties include, in particular, the electric charge, the charge radius,
the anapole moment, and the dipole electric and magnetic moments. Such
nontypical neutrino features are of particular interest, because they open
a door to ``new physics'' beyond the SM (BSM). In spite of appreciable
efforts in searches for neutrino electromagnetic characteristics, up to
now there is no experimental evidence favoring their nonvanishing values.

Among the electromagnetic properties of neutrinos, the most studied and
well understood theoretically are neutrino magnetic moments (NMM), along
with electric dipole moments. For the most recent and complete review on
theoretical and experimental aspects of NMM, as well as for the
corresponding references, see~\cite{Giunti:2014ixa}. The effective
Lagrangian, which describes the coupling of NMM to the electromagnetic
field $F^{\alpha \beta}$, can be written in the form
\begin{equation}\label{Lagr_sigma_F}
L_{int}=\frac{1}{2}\,{\bar \psi}_i \sigma_{\alpha \beta}(\mu
_{ij}+\epsilon_{ij} \gamma_{5}) \psi_j F^{\alpha \beta}+{\rm h.c.},
\end{equation}
where the magnetic moments $\mu_{i j}$, in the presence of mixing between
different neutrino states, are associated with the neutrino mass
eigenstates $\nu_i$. An interplay between the magnetic moment and neutrino
mixing effects is important. Note that the electric (transition) moments
$\epsilon_{ij}$ do also contribute to the coupling. A Dirac neutrino may
have non-zero diagonal electric moments in models where CP invariance is
violated. For a Majorana neutrino the diagonal magnetic and electric
moments are zero. Therefore, NMM can be used to distinguish Dirac and
Majorana neutrinos \cite{nieves82, kayser82, kayser84}.

In the Standard Model the magnetic moment of a massless neutrino is zero.
In the minimal extension of the SM, the explicit evaluation of the
one-loop contributions to the Dirac NMM in the leading approximation over
small parameters $b_i=\frac{m_{i}^{2}}{M_{W}^{2}}$ ($m_i$ are the neutrino
masses, $i=1,2,3$), that however exactly accounts for the parameters $a_l=
\frac{m_{l}^{2}}{M_{W}^{2}}$ ($l=e, \mu, \tau$), yields the following
result~\cite{Petcov:1976ff,fujikawa80,Pal:1981rm,shrock82}:
\begin{equation}\label{m_e_mom_i_j}
  \mu^{D}_{ij}=\frac{e G_F m_{i}}{8\sqrt {2} \pi ^2}
  \left(1 + \frac{m_j}{m_i}\right)\sum_{l= e,  \mu,
  \tau}f(a_l)U_{lj}U^{\ast}_{li},
\end{equation}
where $U_{li}$ is the neutrino mixing matrix, and
\begin{equation}\nonumber
f(a_l)=\frac{3}{4}\left[1+\frac{1}{1-a_l}-\frac{2 a_l}
{(1-a_l)^2}-\frac{2a_l^2\ln a_l}{(1-a_l)^3} \right].
\end{equation}
A Majorana neutrino can also have nondiagonal (transition) magnetic moment
$\mu^{M}_{ij}=2\mu^{D}_{ij}$ ($i\neq j$). The obtained value for NMM is
proportional to the neutrino mass and is, in general, of the order $\sim
10^{-(19\div 21)}\mu_{B}$.

%
Much larger values for NMM can be obtained in different other SM
extensions (see \cite{broggini12,Giunti:2014ixa} for the detailed
discussion). However, there is a problem \cite{Beacom:1999wx} for any BSM
theory how to get a large NMM value and simultaneously to avoid an
unacceptable large contribution to the neutrino mass. Recently this
problem has been reconsidered for a class of BSM theories and it has been
shown in a model-independent way that in principle it is possible to avoid
the above mentioned contradiction in the case of Dirac \cite{Bell:2005kz}
and Majorana \cite{Bell:2006wi} neutrinos. It has been shown that in this
kind of theoretical models the NMM can naturally reach values of $\sim
10^{-(14\div 15)}\mu_{B}$. These values are at least two orders of
magnitude smaller than the present laboratory experimental limits (see
bellow).  There is also a huge gap of many orders of magnitude between
these values and the prediction of the minimal extension of the SM.
Therefore, if any direct experimental confirmation of non-zero NMM is
obtained in the laboratory experiments, it will open a window to ``new
physics''.

The neutrino magnetic moments are being searched in
reactor~\cite{TEXONO,GEMMA}, accelerator~\cite{LSND,DONUT} and
solar~\cite{Super-Kamiokande,Borexino} experiments on low-energy elastic
neutrino-electron scattering (for more details see the review
articles~\cite{broggini12, Giunti:2014ixa} and references therein). The
current best upper limit on the NMM value obtained in such direct
laboratory measurements is
$$
\mu_\nu\leq2.9\times10^{-11}\mu_B,
$$
where $\mu_B=e/(2m_e)$ is a Bohr magneton. This bound, which is due to the
GEMMA experiment~\cite{GEMMA} with a HPGe detector at Kalinin nuclear
power station, is by an order of magnitude larger than the tightest
constraint obtained in astrophysics~\cite{raffelt90}:
$$
\mu_\nu\leq3\times10^{-12}\mu_B.
$$
And it by many orders of magnitude exceeds the value derived in the
minimally extended SM that includes right-handed
neutrinos~\cite{fujikawa80,shrock82}:
$$
\mu_\nu\leq3\times10^{-19}\mu_B \left(\frac{m_\nu}{1\,{\rm eV}}\right),
$$
where $m_\nu$ is a neutrino mass. At the same time, there are different
theoretical BSM scenarios that predict much higher $\mu_\nu$ values. For
example, the effective NMM value in a class of extra-dimension models can
be as large as about $10^{-10}\mu_B$~\cite{mohapatra04}. Future higher
precision reactor experiments can therefore be used to provide new
constraints on large extra-dimensions.

The paper is organized as follows. Sec.~\ref{NMM_searches} outlines the
current status of searches for NMM and the problem of atomic-ionization
effects in reactor experiments. Sec.~\ref{gen_theory} is devoted to the
theoretical background for neutrino scattering on atomic electrons. In
Sec.~\ref{1e}, we discuss the case of neutrino scattering on one bound
electron. Hydrogen-like states and a semiclassical limit are considered.
Sec.~\ref{MEA} focuses on ionization of manyelectron atoms by neutrino
impact. The case of a helium atomic target and the Thomas-Fermi and
\emph{ab initio} approaches are discussed. Finally, Sec.~\ref{summary}
summarizes this review.

\section{Searches for neutrino magnetic moments of reactor antineutrinos}
\label{NMM_searches}
The strategy of experiments searching for NMM is as follows. One studies
an inclusive cross section for elastic (anti)neutrino-electron scattering
which is differential in the energy transfer $T$. In the ultrarelativistic
limit $m_\nu\to0$, it is given by an incoherent sum of the SM contribution
$d\sigma_{SM}/dT$, which is due to weak interaction that conserves the
neutrino helicity, and the helicity-flipping contribution
$d\sigma_{(\mu)}/dT$, which is due to $\mu_\nu$,
\begin{eqnarray}
\label{cr_sec}\frac{d\sigma}{dT}=\frac{d\sigma_{\rm
SM}}{dT}+\frac{d\sigma_{(\mu)}}{dT}.
\end{eqnarray}
The SM term is well-documented and is given by~\cite{cb_book}
\begin{eqnarray}
\label{cr_sec_SM}\frac{d\sigma_{\rm
SM}}{dT}=\frac{G_F^2m_e}{2\pi}\left[(g_V+g_A)^2+(g_V-g_A)^2\left(1-\frac{T}{E_\nu}\right)^2+(g_A^2-g_V^2)\frac{m_eT}{E_\nu^2}\right],
\end{eqnarray}
where $E_\nu$ is the incident antineutrino energy, $g_A=1/2$ and
$g_V=(4\sin^2\theta_W+1)/2$ for $\nu_e$ and $g_A=-1/2$ and
$g_V=(4\sin^2\theta_W-1)/2$ for $\nu_\mu$ and $\nu_\tau$, with $\theta_W$
being the Weinberg angle. For antineutrinos one must substitute
$g_A\to-g_A$.

The possibility for neutrino-electron elastic scattering due to NMM was
first considered in Ref.~\cite{carlson32}, and the cross section of this
process was calculated in Ref.~\cite{bethe35} (the related brief
historical notes can be found in Ref.~\cite{kyuldjiev84}). Here we would
like to recall the paper by Domogatsky and Nadezhin~\cite{domogatskii70},
where the cross section of Ref.~\cite{bethe35} was corrected and the
antineutrino-electron cross section was considered in the context of the
earlier experiments with reactor antineutrinos of Cowan and
Reines~\cite{cowan57} and Cowan et al.~\cite{cowan54}, which were aimed to
reveal the NMM effects. Discussions on the derivation of the cross section
and on the optimal conditions for bounding the NMM value, as well as a
collection of the cross section formulas for elastic scattering of
neutrinos (antineutrinos) on electrons, nucleons, and nuclei, can be found
in Refs.~\cite{kyuldjiev84} and~\cite{vogel89}. The result relevant to the
$\mu_\nu$ component in Eq.~(\ref{cr_sec})
reads~\cite{kyuldjiev84,domogatskii70,vogel89}
\begin{eqnarray}
\label{cr_sec_mu}\frac{d\sigma_{(\mu)}}{dT}=4\pi\alpha\mu_\nu^2\left(\frac{1}{T}-\frac{1}{E_\nu}\right),
\end{eqnarray}
where $\alpha$ is the fine-structure constant. Thus, the two components of
the cross section~(\ref{cr_sec}) exhibit qualitatively different
dependencies on the recoil-electron kinetic energy $T$. Namely, at low $T$
values the SM cross section is practically constant in $T$, while that due
to $\mu_\nu$ behaves as $1/T$. This means that the experimental
sensitivity to NMM value critically depends on lowering the energy
threshold of the detector employed for measurement of the recoil-electron
spectrum.

The current reactor experiments with germanium
detectors~\cite{TEXONO,GEMMA} have reached threshold values of $T$ as low
as few keV and are to further improve the sensitivity to low energy
deposition in the detector~\cite{wong11,li13,li14}. At low energies,
however, one can expect a modification of the free-electron formulas due
to the binding of electrons in the germanium atoms, where e.g. the energy
of the $K_\alpha$ line, 9.89\,keV, indicates that at least some of the
atomic binding energies are comparable to the already relevant to the
experiment values of $T$. Thus a proper treatment of the atomic effects in
neutrino scattering is necessary and important for the analysis of the
current and, even more, of the future data with a still lower threshold.
Furthermore, there is no known means of independently calibrating
experimentally the response of atomic systems, such as the germanium, to
the scattering due to the interactions relevant for the neutrino
experiments. Therefore one has to rely on a pure theoretical analysis in
interpreting the neutrino data. For the first time this problem was
addressed in Ref.~\cite{gdt75}, where a 2-3 times enhancement of the
electroweak cross section in the case of ionization from a 1$s$ state of a
hydrogen-like atom with nuclear charge $Z$ had been numerically determined
at neutrino energies $E_\nu\sim\alpha Zm_ec^2$. Subsequent numerical
calculations within the relativistic Hartree-Fock method for ionization
from inner shells of various atoms showed much lower enhancement
($\sim5-10\%$) of the electroweak
contribution~\cite{ddf92,fdd92,kmsf,fms,gpp02,kms03}. It was found that in
the scattering on realistic atoms, such as germanium, the so-called
stepping approximation works with a very good accuracy. The stepping
approach, introduced in Ref.~\cite{kmsf} from an interpretation of
numerical data, treats the process as scattering on individual independent
electrons occupying atomic orbitals and suggests that the cross section
follows the free-electron behavior in Eqs.~(\ref{cr_sec_SM}) and
(\ref{cr_sec_mu}) down to $T$ equal to the ionization threshold for the
orbital, and that below that energy the electron on the corresponding
orbital is `deactivated' thus producing a sharp `step' in the dependence
of the cross section on $T$.

The interest to the role of atomic effects was renewed in several more
recent papers. The early claim~\cite{wll} of a significant enhancement of
the NMM contribution in the case of germanium due to the atomic effects
has been later disproved~\cite{mv,wll2} and it was
argued~\cite{ks,npb11,jpl11,prd11} that the modification of the
free-electron formulas (\ref{cr_sec_SM}) and (\ref{cr_sec_mu}) by the
atomic-binding effects is insignificant down to very low values of $T$.
This conclusion appeared to be also in contradiction with the results of
Ref.~\cite{martemyanov11}, where it was deduced by means of numerical
calculations that the $\mu_\nu$ contribution to ionization of the helium
atomic target by impact of electron antineutrinos strongly enhances
relative to the free-electron approximation. However, from calculations
performed in Ref.~\cite{pepan_lett14} it follows that the stepping
approximation is well applicable practically down to the ionization
threshold for helium.

%
\section{General theoretical framework}
\label{gen_theory}
As indicated in the introduction, the most sensitive and widely used
method for the experimental investigation of the neutrino electromagnetic
properties is provided by direct laboratory measurements of low-energy
elastic scattering of neutrinos and antineutrinos with electrons in
reactor, accelerator and solar experiments. In this section we deliver a
theoretical background for such studies.

\subsection{Neutrino-electron interactions}
\label{nu-e}
Let us consider the elastic-scattering process
\begin{equation}
\label{nu-e_process} \nu+e^-\to\nu+e^-,
\end{equation}
where an incident neutrino with energy $E_\nu$ transfers to a free
electron, which is initially at rest in the laboratory frame, the
energy-momentum $q$. There are two recoil-electron observables: the
kinetic energy $T$, which amounts to the energy transfer, and the outgoing
angle $\chi$ measured with respect to the incident neutrino direction. In
the ultrarelativistic limit $m_\nu=0$, these kinematical variables are
related by
\begin{equation}
\label{chi-to-T} \cos\chi=\frac{E_\nu+m_e}{E_\nu}\sqrt{\frac{T}{T+2m_e}}.
\end{equation}
The maximal value of the kinetic electron energy is thus realized when
$\chi=0^\circ$ and is given by
\begin{equation}
\label{T_max} T_{max}=\frac{2E_\nu^2}{2E_\nu+m_e}.
\end{equation}

\begin{figure}
\begin{center}
\includegraphics[width=0.45\textwidth]{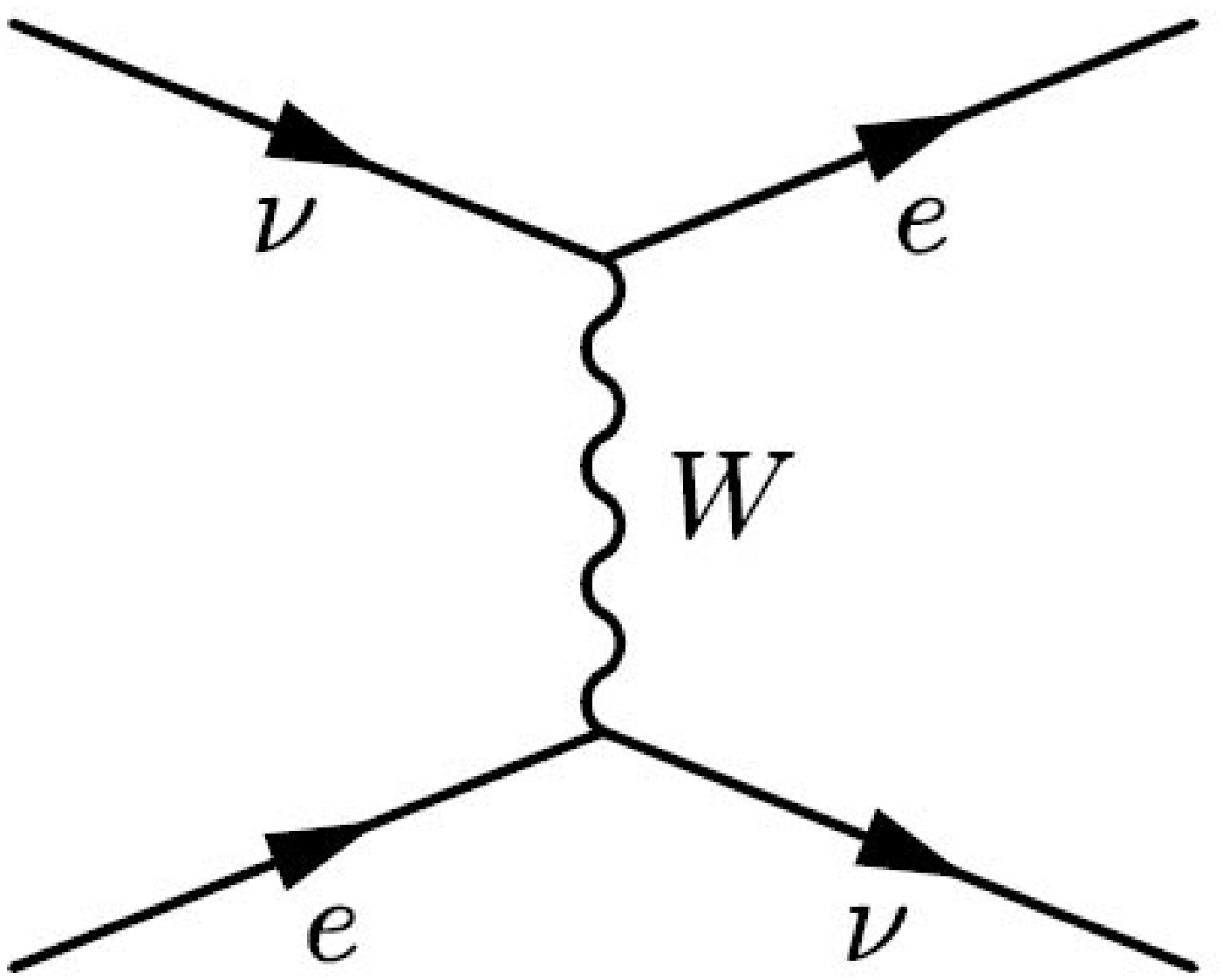}
\includegraphics[width=0.45\textwidth]{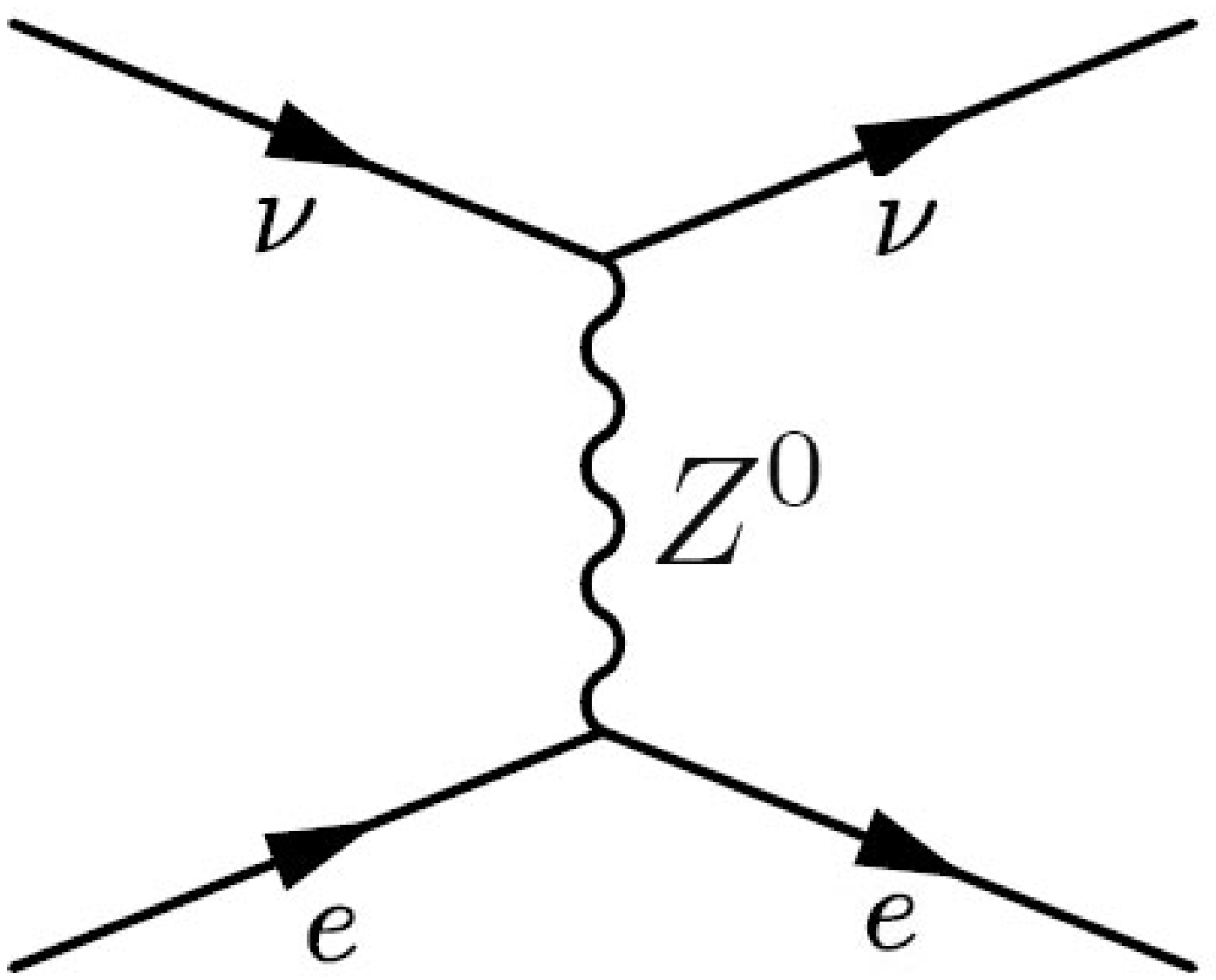}
\end{center}
\caption{\label{fig:e-nu_SM} Elastic neutrino-electron scattering due to
the weak interaction. Exchange by the $W$ (left) and $Z^0$ (right) bosons
is shown.}
\end{figure}

Within the SM, the scattering process~(\ref{nu-e_process}) takes place due
to exchange of the weak bosons, as shown in Fig.~\ref{fig:e-nu_SM}.
The $W$-boson channel corresponds to the charged current interaction and
is absent in the cases of the muon and tau neutrinos. If $|q^2|\ll m_W^2$,
where $m_W$ is the $W$-boson mass, the scattering amplitude is given
by~\cite{cb_book}
\begin{equation}
\label{M_W} M_{W}=\frac{G_F}{\sqrt{2}}\bar{u}_{\nu_{2}}
\gamma_\alpha(1-\gamma_5) u_{\nu_{1}}
\bar{u}_{e_{2}}\gamma^\alpha(1-\gamma_5)u_{e_{1}},
\end{equation}
where $u_{\nu_{1}}$ ($u_{e_1}$) and $u_{\nu_{2}}$ ($u_{e_2}$) are initial
and final neutrino (electron) spinors. The $Z^0$ boson mediates the
neutral current interaction. The corresponding scattering amplitude in the
case $|q^2|\ll m_Z^2$, where $m_Z$ is the $Z^0$-boson mass,
reads~\cite{cb_book}
\begin{equation}
\label{M_Z} M_{Z}=\frac{G_F}{\sqrt{2}}\bar{u}_{\nu_{2}}
\gamma_\alpha(1-\gamma_5) u_{\nu_{1}}
\bar{u}_{e_{2}}\gamma^\alpha(g_V-g_A\gamma_5)u_{e_{1}}.
\end{equation}
Using the matrix elements (\ref{M_W}) and (\ref{M_Z}), one arrives, after
averaging over the initial and summing over the final electron spins, at
the SM single-differential cross section~(\ref{cr_sec_SM}).

\begin{figure}
\begin{center}
\includegraphics[width=0.45\textwidth]{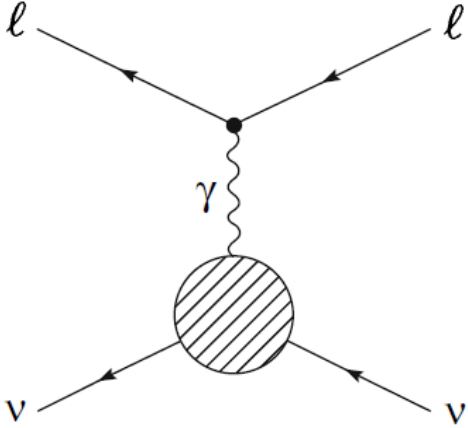}
\end{center}
\caption{\label{fig:e-nu_EM} Contribution of the neutrino electromagnetic
vertex function to neutrino elastic scattering on a charged
lepton~\cite{giunti09}.}
\end{figure}

Fig.~\ref{fig:e-nu_EM} shows the electromagnetic channel of the scattering
process~(\ref{nu-e_process}).
In general the matrix element of the neutrino electromagnetic current
$J_\mu$ can be considered between different neutrino initial $\psi_i(p)$
and final $\psi_j(p')$ states of different masses, $p^2=m^2_i$ and
$p'^2=m^2_j$:
\begin{equation}
\label{matrix_el}\langle\psi_j(p')|J_\mu|\psi_i(p)\rangle=\bar{u}_j(p')\Lambda_\mu(q)u_i(p).
\end{equation}
In the most general case consistent with Lorentz and electromagnetic gauge
invariance, the vertex function is defined as (see
Refs.~\cite{giunti09,broggini12} and references therein)
\begin{equation}
\label{vertex}\Lambda_\mu(q)=[f_Q(q^2)_{ij} +
f_A(q^2)_{ij}\gamma_5](q^2\gamma_\mu-\gamma_\mu\!\not{\!q})+f_M(q^2)_{ij}i\sigma_{\mu\nu}q^\nu+f_E(q^2)_{ij}i\sigma_{\mu\nu}q^\nu\gamma_5,
\end{equation}
where $f_Q(q^2)$, $f_A(q^2)$, $f_M(q^2)$, and $f_E(q^2)$ are respectively
the charge, anapole, dipole magnetic, and dipole electric neutrino form
factors, which are matrices in the space of neutrino mass
eigenstates~\cite{shrock82}.

Consider the diagonal case $i=j$. The hermiticity of the electromagnetic
current and the assumption of its invariance under discrete symmetries'
transformations put certain constraints on the form factors, which are in
general different for the Dirac and Majorana neutrinos. In the case of
Dirac neutrinos, the assumption of CP invariance combined with the
hermiticity of the electromagnetic current $J_\mu$ implies that the
electric dipole form factor vanishes, $f_E=0$. At zero momentum transfer
only $f_Q(0)$ and $f_M(0)$, which are called the electric charge and the
magnetic moment, respectively, contribute to the Hamiltonian $H_{\rm
int}\sim J_\mu A^\mu$ that describes the neutrino interaction with the
external electromagnetic field $A^\mu$. The hermiticity also implies that
$f_Q$, $f_A$, and $f_M$ are real. In contrast, in the case of Majorana
neutrinos, regardless of whether CP invariance is violated or not, the
charge, dipole magnetic and electric moments vanish, $f_Q=f_M=f_E=0$, so
that only the anapole moment can be non-vanishing among the
electromagnetic moments. Note that it is possible to
prove~\cite{nieves82,kayser82,kayser84} that the existence of a
non-vanishing magnetic moment for a Majorana neutrino would bring about a
clear evidence for CPT violation.

In the off-diagonal case $i\neq j$, the hermiticity by itself does not
imply restrictions on the form factors of Dirac neutrinos. It is possible
to show~\cite{nieves82} that, if the assumption of the CP invariance is
added, the form factors $f_Q$, $f_M$, $f_E$, and $f_A$ should have the
same complex phase. For the Majorana neutrino, if CP invariance holds,
there could be either a transition magnetic or a transition electric
moment. Finally, as in the diagonal case, the anapole form factor of a
Majorana neutrino can be nonzero.

The neutrino dipole magnetic and electric form factors (and the
corresponding magnetic and electric dipole moments) are theoretically the
most well understood among the form factors. The value of the magnetic
form factor $f_M(q^2)$ at $q^2=0$ defines the NMM, $\mu_\nu=f_M(0)$. In
the low-energy limit, the NMM contribution to the effective
electromagnetic vertex can be expressed in the following form:
\begin{equation}
\label{el-m_vertex}\Lambda_{\alpha}=\frac{\mu_{\nu}}{2m_{e}}\,\sigma_{\alpha\beta}q^{\beta}.
\end{equation}
Thus, the corresponding scattering amplitude is
\begin{equation}
M_{(\mu)}=\frac{4\pi \mu_\nu \sqrt{\alpha}}{2m_eq^2} \, \bar{u}_{\nu_{2}}
\sigma^{\alpha \beta} q_\beta u_{\nu_{1}} \bar{u}_{e_{2}} \gamma_\alpha
u_{e_{1}}.
\end{equation}
This leads to the NMM single-differential cross section given
by~(\ref{cr_sec_mu}).

\subsection{Neutrino scattering on atomic electrons}
\label{nu-at_el}
Consider the process where a neutrino with energy-momentum
$p_{\nu}=(E_{\nu}, {\bf p}_{\nu})$ scatters on an atom at energy-momentum
transfer $q=(T,{\bf q})$. In what follows the recoil of the atomic nucleus
is neglected because of the typical for current experiments situation
$T\gg 2E_{\nu} / M_N$, where $M_N$ is the nuclear mass. The atomic target
is supposed to be unpolarized and in its ground state ${\mid}0\rangle$
with the corresponding energy $E_0$. It is also supposed that $T\ll m_{e}$
and $\alpha Z \ll 1$, where $Z$ is the nuclear charge and $\alpha$ is the
fine-structure constant, so that the initial and final electronic systems
can be treated nonrelativistically. The neutrino states are described by
the Dirac spinors assuming $m_{\nu}=0$.

Thus, the magnetic moment interaction of the neutrino field $\psi$ with
the atomic electrons is described by the Lagrangian
\begin{equation}\label{Lagr_NuMM}
L_{int}=\frac{\mu_{\nu}}{2m_{e}}\,{\bar \psi}(p'_\nu)\sigma_{\alpha
\beta}\psi (p_\nu) q^{\alpha} A^{\beta},
\end{equation}
where $p_\nu'$ is the final neutrino four-momentum. The electromagnetic
field $A=(A_0,{\bf A})$ of the atomic electrons is\footnote{Hereafter we
use the notation $q=|{\bf q}|$.} $A_0({\bf q})=\sqrt{4 \pi \alpha} \,
\rho( {\bf q})/q^2$, ${\bf A}({\bf q})=\sqrt{4 \pi \alpha} \,{\bf j}( {\bf
q})/q^2$, where $\rho({\bf q})$ and ${\bf j}({\bf q})$ are the Fourier
transforms of the electron number density and current density operators,
respectively,
\begin{equation} \rho({\bf q})= \sum_{a=1}^Z \exp(i {\bf q} \cdot {\bf r}_a),
\label{ne}
\end{equation}
\begin{equation} {\bf j}({\bf q})= -\frac{i}{2m}\sum_{a=1}^Z \left[\exp(i
{\bf q} \cdot {\bf r}_a)\frac{\partial}{\partial{\bf
r}_a}+\frac{\partial}{\partial{\bf r}_a}\exp(i {\bf q} \cdot {\bf
r}_a)\right], \label{ce}
\end{equation}
and the sums run over the positions ${\bf r}_a$ of all the $Z$ electrons
in the atom. The double-differential cross section can be presented as
\begin{equation}
\label{dcs}
\frac{d^{2}\sigma_{(\mu)}}{dTdq^{2}}=\left(\frac{d^{2}\sigma_{(\mu)}}{dTd{q}^{2}}\right)_\parallel+
\left(\frac{d^{2}\sigma_{(\mu)}}{dTd{q}^{2}}\right)_\perp,
\end{equation}
where
\begin{equation} \left({d^{2} \sigma_{(\mu)} \over dT \, d {q}^{2}}\right)_\parallel = 4 \pi \, \alpha \, { \mu_{\nu}^{2} \over {q}^{2}}
\left(1-\frac{T^{2}}{{q}^{2}}\right)S(T,{q}^{2}), \label{dcs_par}
\end{equation}
and
\begin{equation} \left({d^{2} \sigma_{(\mu)} \over dT \, d {q}^{2}}\right)_\perp = 4 \pi \, \alpha \, { \mu_{\nu}^{2} \over {q}^{2}} \,
\left(1-\frac{{q}^{2}}{4E_{\nu}^{2}}\right)R(T,{q}^{2}), \label{dcs_perp}
\end{equation}
where $S(T,{q}^{2})$, also known as the dynamical structure
factor~\cite{vh}, and $R(T,{q}^{2})$ are
\begin{equation}S(T,{q}^{2})=\sum_n \, \delta (T - E_n+E_0) \, \left | \langle n |
\rho({\bf q}) | 0 \rangle \right |^{2}, \label{dsf}
\end{equation}
\begin{equation} R(T,{q}^{2})=\sum_n \, \delta (T - E_n+E_0) \, \left | \langle n |
j_\perp({\bf q}) | 0 \rangle \right |^{2}, \label{dsfc}
\end{equation}
with $j_\perp$ being the ${\bf j}$ component perpendicular to ${\bf q}$
and parallel to the scattering plane, which is formed by the incident and
final neutrino momenta. The sums in Eqs.~(\ref{dsf}) and~(\ref{dsfc}) run
over all the atomic states $| n \rangle$ with energies $E_n$ of the
electron system, with $|0 \rangle$ being the initial state.

The longitudinal term~(\ref{dcs_par}) is associated with atomic
excitations induced by the force that the neutrino magnetic moment exerts
on electrons in the direction parallel to ${\bf q}$. The transverse
term~(\ref{dcs_perp}) corresponds to the exchange of a virtual photon
which is polarized as a real one, that is, perpendicular to ${\bf q}$. It
resembles a photoabsorption process when $q \to T$ and the virtual-photon
four-momentum thus approaches a real-photon value. Due to selections
rules, the longitudinal and transverse excitations do not interfere (see
\cite{fano63} for detail).

The factors $S(T,{q}^{2})$ and $R(T,{q}^{2})$ are related to respectively
the density-density (or polarization) and current-current Green's
functions
\begin{equation} S(T,{q}^{2})={1 \over \pi} \, {\rm Im}\,F(T,{q}^{2}), \label{sfrel}
\end{equation}
\begin{equation} R(T,{q}^{2})={1 \over \pi} \, {\rm Im}\,L(T,{q}^{2}), \label{sfrelc}
\end{equation}
where
\begin{equation}F(T,{q}^{2})=\sum_n {\left | \langle n | \rho({\bf q}) | 0 \rangle \right
|^{2} \over T - E_n+E_0 - i \, \epsilon} \,  = \left \langle 0 \left
|\rho(- {\bf q}) \, {1 \over T-H+E_0- i \, \epsilon}\, \rho({\bf q})
\right | 0 \right \rangle, \label{fdef}
\end{equation}
\begin{equation} L(T,{q}^{2})=\sum_n {\left | \langle n | j_\perp({\bf q)} | 0 \rangle
\right |^{2} \over T - E_n+E_0 - i \, \epsilon} \,  = \left \langle 0
\left |j_\perp(- {\bf q}) \, {1 \over T-H+E_0- i \, \epsilon}\,
j_\perp({\bf q}) \right | 0 \right \rangle, \label{fdefc}
\end{equation}
$H$ being the Hamiltonian for the system of electrons. From these
relations it follows that, due to the parity selection rule, the functions
$S(T,{q}^{2})$ and $R(T,{q}^{2})$ are even with respect to $q$.

For small ${q}$ values, in particular, such that $q\sim T$, only the
lowest-order non-zero terms of the expansion of Eqs.~(\ref{sfrel})
and~(\ref{sfrelc}) in powers of ${q}^{2}$ are of relevance (the so-called
dipole approximation). In this case, one has~\cite{mv,ks}
\begin{equation}R(T,{q}^{2})=\frac{T^{2}}{{q}^{2}}S(T,{q}^{2}). \label{dipole_approx}
\end{equation}
Taking into account Eq.~(\ref{dipole_approx}), the experimentally measured
singe-differential inclusive cross section is, to a good approximation,
given by (see e.g. in \cite{ks,jpl11,prd11})
\begin{equation} {d \sigma_{(\mu)} \over dT } = 4 \pi \, \alpha \, \mu_{\nu}^{2}
\int_{T^{2}}^{(2E_{\nu}-T)^{2}} \, S(T,{q}^{2})\, {d{q}^{2} \over
{q}^{2}}. \label{d1s} \end{equation}

The standard electroweak contribution to the cross section can be
similarly expressed in terms of the same factor
$S(T,{q}^{2})$~\cite{mv,prd11} as
\begin{eqnarray}{d \sigma_{SM} \over dT } &=& {G_F^{2} \over 4 \pi} \left ( 1+ 4\,\sin^{2} \theta_{W} + 8 \, \sin^4 \theta_{W} \right
)\nonumber\\&{}&\times \int_{T^{2}}^{(2E_{\nu}-T)^{2}} \, S(T,{q}^{2}) \,
d{q}^{2}, \label{d1sw}
\end{eqnarray}
where the factor $S(T,{q}^{2})$ is integrated over ${q}^{2}$ with a unit
weight, rather than ${q}^{-2}$ as in Eq.~(\ref{d1s}).

The kinematical limits for ${q}^{2}$ in an actual neutrino scattering are
explicitly indicated in Eqs.~(\ref{d1s}) and (\ref{d1sw}). At large
$E_{\nu}$, typical for the reactor neutrinos, the upper limit can in fact
be extended to infinity, since in the discussed here nonrelativistic limit
the range of momenta $\sim E_{\nu}$ is indistinguishable from infinity on
an atomic scale. The lower limit can be shifted to ${q}^{2}=0$, since the
contribution of the region of ${q}^{2} < T^{2}$ can be expressed in terms
of the photoelectric cross section~\cite{mv} and is negligibly small (at
the level of below one percent in the considered range of $T$). For this
reason one can discuss the momentum-transfer integrals in Eqs.~(\ref{d1s})
and~(\ref{d1sw}) running from ${q}^{2}=0$ to ${q}^{2}=\infty$:
\begin{equation} I_1(T)=\int_0^{\infty} \, S(T,{q}^{2})\, {d{q}^{2} \over
{q}^{2}}\label{defi_1},\end{equation}~
 and
 \begin{equation} I_2(T)=\int_0^{\infty} \, S(T,{q}^{2})\, d{q}^{2}~. \label{defi_2}\end{equation}

For a free electron, which is initially at rest, the
density-density correlator is the free particle Green's function
\begin{equation} F_{(FE)}(T,{q}^{2})= \left ( T-{{q}^{2} \over 2m_e} - i \,
\epsilon \right  )^{-1}, \label{ff} \end{equation}
so that the dynamical structure factor is given by
\begin{equation}
\label{struc_factor_FE}
S_{(FE)}(T,{q}^{2})=\delta\left(T-\frac{{q}^{2}}{2m_e}\right),
\end{equation}
 and the discussed here
integrals are in the free-electron limit as follows:
\begin{equation} I_1^{(FE)}=\int_0^{\infty} \, S_{(FE)}(T,{q}^{2})\, {d{q}^{2} \over {q}^{2}} = {1 \over T},\label{intf_1}
\end{equation}
\begin{equation} I_2^{(FE)}=\int_0^{\infty} \, S_{(FE)}(T,{q}^{2})\, d{q}^{2} = 2 \, m_e. \label{intf_2}
\end{equation}
Clearly, these expressions, when used in the formulas (\ref{d1s}) and
(\ref{d1sw}), result in the free-electron cross sections for the case
$T\ll E_\nu$,
\begin{equation} {d \sigma_{(\mu)} \over dT }= \frac{4 \pi \, \alpha \,
\mu_{\nu}^{2}}{T} \label{fe}
\end{equation}
 and
\begin{eqnarray} {d \sigma_{SM} \over dT }= {G_F^{2} \, m_e \over 2
\pi} \left ( 1+ 4 \, \sin^{2} \theta_{W} + 8 \, \sin^4 \theta_{W} \right
),
\label{sew}
\end{eqnarray}
 correspondingly.

\section{Scattering on one bound electron}
\label{1e}
In this section we consider neutrino scattering on an electron bound in an
atom following consideration of \cite{ks,jpl11,prd11}. The binding effects
generally deform the density-density Green's function, so that both the
integrals (\ref{defi_1}) and (\ref{defi_2}) are somewhat modified. Namely,
the binding effects spread the free-electron $\delta$-peak in the
dynamical structure function~(\ref{struc_factor_FE}) at ${q}^{2}=2 m_e T$
and also shift it by the scale of characteristic electron momenta in the
bound state.

\subsection{Ionization from a hydrogen-like orbital} \label{H}
Consider the situation when the initial electron occupies the discrete
$nl$ orbital in a Coulomb potential $V({\bf r})=-\alpha Z/r$. The
dynamical structure factor for this hydrogen-like system is given by
\begin{equation} S_{(nl)}(T,q^2)= \frac{m_ek}{(2\pi)^3}\frac{1}{2l+1}\sum_{m=-l}^l
\int d\Omega_k|\langle\varphi^-_{{\bf k}}|\rho({\bf
q})|\varphi_{nlm}\rangle|^2, \label{sfnl}
\end{equation}
where $\varphi_{nlm}$ is the bound-state wave function, $\varphi^-_{{\bf
k}}$ is the outgoing Coulomb wave for the ejected electron with momentum
${\bf k}$, and $k=|{\bf k}|=\sqrt{2m_eT-p_n^2}$, with $p_n=\alpha Zm_e/n$
being the electron momentum in the $n$th Bohr orbit. The closed-form
expressions for the bound-free transition matrix elements in
Eq.~(\ref{sfnl}) can be found, for instance, in Ref.~\cite{belckic81}. In
principle, they allow for performing angular integrations in
Eq.~(\ref{sfnl}) analytically. This task, however, turns out to be
formidable for large values of $n$. Therefore, below we restrict our
consideration to the $n=1,2$ states only, which nevertheless is enough for
demonstrating the validity of the semiclassical approach developed in
Sec.~\ref{1e}.

Using results of Ref.~\cite{holt69}, we can present the
function~(\ref{sfnl}) when $n=1,2$ as
\begin{eqnarray}
\label{sfnl1} S_{(nl)}(T,q^2)&=&\frac{2^8m_ep_n^6}{3[1-\exp(-2\pi\eta)]}\frac{q^2f_{nl}(q^2)}{[(q^2-k^2+p_n^2)^2+4p_n^2k^2]^{2n+1}}\nonumber\\
&{}&\times
\exp\left[-2\eta\arctan\left(\frac{2p_nk}{q^2-k^2+p_n^2}\right)\right ] ,
\end{eqnarray}
where the branch of the arctangent function should be used that lies
between 0 and $\pi$, $\eta=\alpha Zm_e/k$ is the Sommerfeld parameter, and
\begin{equation}
\label{f1s} f_{1s}(q^2)=3q^2+k^2+p_1^2,
\end{equation}
\begin{eqnarray}
\label{f2s} f_{2s}(q^2)&=&8\Bigg[3q^{10}-(32p_2^2+11k^2)q^8+(82p_2^4+72p_2^2k^2+14k^2)q^6\nonumber\\
&{}&\left.+(20p_2^6-62p_2^4k^2-20p_2^2k^4-6k^6)q^4+(p_2^2+k^2)\right.\nonumber\\
&{}&\left.\times\left(\frac{47}{5}p_2^6-\frac{47}{5}p_2^4k^2-7p_2^2k^4-k^6\right)q^2+(4p_2^2+k^2)(p_2^2+k^2)^4\right],\nonumber\\
\end{eqnarray}
\begin{eqnarray}
\label{f2p} f_{2p}(q^2)&=&2p_2^2\Bigg[36q^8-48(p_2^2+k^2)q^6+(152p_2^4-48p_2^2k^2-8k^4)q^4+(p_2^2+k^2)\nonumber\\
&{}&\left.\times\left(\frac{1712}{15}p_2^4+\frac{1568}{15}p_2^2k^2+16k^4\right)q^2
+\left(\frac{44}{3}p_2^2+4k^2\right)(p_2^2+k^2)^3\right].\nonumber\\
\end{eqnarray}

\begin{figure}
\begin{center}
\includegraphics[width=0.9\textwidth]{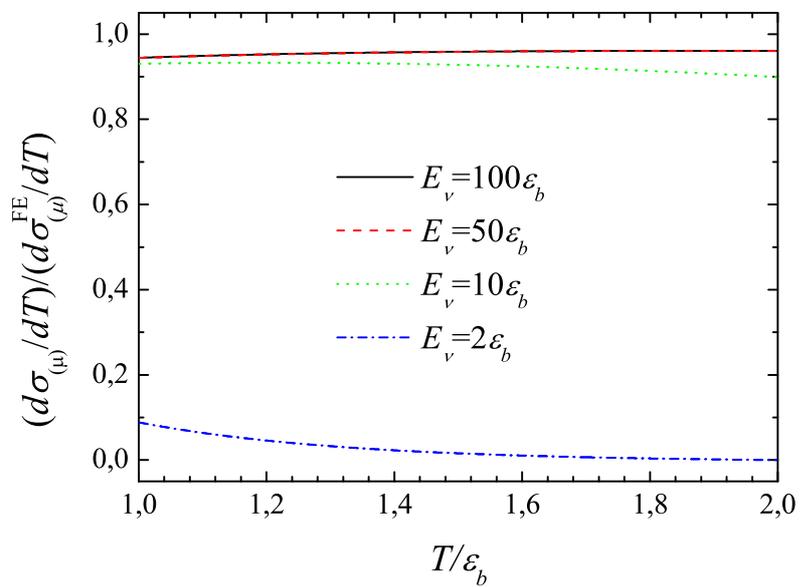}
\end{center}
\caption{\label{fig1} The ratio of single-differential cross sections for
magnetic neutrino scattering from the $1s$ hydrogen-like and free-electron
states, respectively, versus $T/\varepsilon_b$ at different values of
$E_\nu$. The $E_\nu=50\,\varepsilon_b$ and $E_\nu=100\,\varepsilon_b$
curves are practically indistinguishable.}
\end{figure}

Fig.~\ref{fig1} shows the magnetic single-differential cross
section~(\ref{d1s}) for ionization from the $1s$ orbital, which is
normalized to the free-electron value~(\ref{cr_sec_mu}), as a function of
$T/\varepsilon_b$, with the electron binding energy given by
$\varepsilon_b=\alpha^2Z^2m_e/2$. As can be seen, the numerical results
for $E_\nu\gg\varepsilon_b$ are close to the free-electron ones in
magnitude. This can be qualitatively explained by noticing the following
facts. First, in an attractive Coulomb potential there is an infinite set
of bound states, with the discrete spectrum smoothly transforming into the
continuum at the ionization threshold. Second, the average value of the
$1s$-electron momentum is $p_e=p_1$ and the average change in the electron
momentum after ejection, $\Delta p_e$, is such that $\Delta p_e^2=2m_eT$,
which is analogous to the free-electron case.

Thus, taking into account the results in Fig.~\ref{fig1}, one might expect
the atomic-binding effects to play a subsidiary role when
$E_\nu\gg\varepsilon_b$. The authors of \cite{wll}, however, came to the
contrary conclusion that the single-differential cross section
dramatically enhances due to atomic ionization when $T\sim\varepsilon_b$.
The enhancement mechanism proposed in~\cite{wll} is based on an analogy
with the photoionization process. As mentioned above, when $q\to T$ the
virtual-photon momentum approaches the physical regime $T^2-q^{2}=0$. In
this limit, we have for the transverse component of the
double-differential cross section~(\ref{dcs_perp})
\begin{equation}
\label{photoeffect}\left({d^{2} \sigma_{(\mu)} \over dT \, d
{q}^{2}}\right)_\perp=\frac{\mu_\nu^2}{\pi}\frac{\sigma_\gamma(T)}{T},
\end{equation}
where $\sigma_\gamma(T)$ is the photoionization cross section at the
photon energy $T$~\cite{akhiezer_book}. The limiting
form~(\ref{photoeffect}) was used in \cite{wll} in the whole integration
interval, when deriving the single-differential cross section. Such a
procedure is obviously incorrect, for the integrand rapidly falls down as
${q}^{2}$ ranges from $T^{2}$ up to $(2E_{\nu}-T)^{2}$, especially when
${q}^{2}\gtrsim r_a^{-2}$, where $r_a$ is a characteristic atomic size
(within the Thomas-Fermi model $r_a^{-1}=Z^{1/3}\alpha m_{e}$~\cite{ll}).
This fact reflects a strong departure from the real-photon regime. For
this reason we can classify the enhancement of the differential cross
section determined in \cite{wll} as spurious.

Insertion of Eq.~(\ref{sfnl1}) into the integrals~(\ref{defi_1})
and~(\ref{defi_2}) and integration over $q^2$, using the change of
variable
$$
\frac{2p_nk}{q^2-k^2+p_n^2}=\tan x
$$
and the standard integrals involving the products of the exponential
function and the powers of sine and cosine functions, yields~\cite{prd11}
\begin{eqnarray}
\label{mw1s}I_1^{(1s)}(T)=\frac{I_2^{(1s)}(T)}{2m_eT}&=&\frac{T^{-1}}{1-\exp(-\frac{2\pi}{\sqrt{y_1-1}})}\left\{1-
\exp\left(-\frac{\pi}{\sqrt{y_1-1}}\right)\right.
\nonumber\\&{}&\times\exp\left [
\frac{-2}{\sqrt{y_1-1}}\arctan\left(\frac{y_1-2}{2\sqrt{y_1-1}}\right)\right
]\nonumber\\&{}&\times\left.
\left(1-\frac{4}{y_1}+\frac{16}{3y_1^2}\right)\right \},
\end{eqnarray}
\begin{eqnarray}
\label{m2s}I_1^{(2s)}(T)&=&\frac{T^{-1}}{1-\exp(-\frac{4\pi}{\sqrt{y_2-1}})}\left
\{ 1- \exp\left(-\frac{2\pi}{\sqrt{y_2-1}}\right)\right.
\nonumber\\&{}&\times\exp\left [
\frac{-4}{\sqrt{y_2-1}}\arctan\left(\frac{y_2-2}{2\sqrt{y_2-1}}\right)\right
]\nonumber\\&{}&\left.\times
\left(1-\frac{8}{y_2}+\frac{80}{3y_2^2}-\frac{448}{15y_2^3}+\frac{1792}{15y_2^4}\right)\right \},
\end{eqnarray}
\begin{eqnarray}
\label{w2s}I_2^{(2s)}(T)&=&\frac{2m_e}{1-\exp(-\frac{4\pi}{\sqrt{y_2-1}})}\left
\{ 1- \exp\left(-\frac{2\pi}{\sqrt{y_2-1}}\right)\right.
\nonumber\\&{}&\times\exp\left [
\frac{-4}{\sqrt{y_2-1}}\arctan\left(\frac{y_2-2}{2\sqrt{y_2-1}}\right)\right]\nonumber\\&{}&\left.\times
\left(1-\frac{8}{y_2}+\frac{80}{3y_2^2}-\frac{448}{15y_2^3}+\frac{1024}{15y_2^4}\right)\right \},
\end{eqnarray}
\begin{eqnarray}
\label{m2p}I_1^{(2p)}(T)&=&\frac{T^{-1}}{1-\exp(-\frac{4\pi}{\sqrt{y_2-1}})}\left
\{ 1- \exp\left(-\frac{2\pi}{\sqrt{y_2-1}}\right)\right.
\nonumber\\&{}&\times\exp\left [
\frac{-4}{\sqrt{y_2-1}}\arctan\left(\frac{y_2-2}{2\sqrt{y_2-1}}\right)\right
]\nonumber\\&{}&\left.\times
\left(1-\frac{8}{y_2}+\frac{80}{3y_2^2}-\frac{704}{15y_2^3}+\frac{3328}{45y_2^4}\right)\right \},
\end{eqnarray}
\begin{eqnarray}
\label{w2p}I_2^{(2p)}(T)&=&\frac{2m_e}{1-\exp(-\frac{4\pi}{\sqrt{y_2-1}})}\left
\{ 1- \exp\left(-\frac{2\pi}{\sqrt{y_2-1}}\right)\right.
\nonumber\\&{}&\times\exp\left [
\frac{-4}{\sqrt{y_2-1}}\arctan\left(\frac{y_2-2}{2\sqrt{y_2-1}}\right)\right
]\nonumber\\&{}&\left.\times
\left(1-\frac{8}{y_2}+\frac{80}{3y_2^2}-\frac{704}{15y_2^3}+\frac{512}{15y_2^4}\right)\right \} ,
\end{eqnarray}
where $y_n=2m_eT/p_n^2\equiv T/|E_n|$. The largest deviations of these
integrals from the free-electron analogs~(\ref{intf_1}) and~(\ref{intf_2})
occur at the ionization threshold $T=|E_n|$. The corresponding relative
values in this specific case are~\cite{prd11}
\begin{equation} \label{mw1s_num}
\frac{I^{(1s)}_1}{I^{(FE)}_1}=\frac{I^{(1s)}_2}{I^{(FE)}_2}=1-\frac{7}{3}e^{-4}=0.9572635093,
\end{equation}
$$
\frac{I^{(2s)}_1}{I^{(FE)}_1}=1-\frac{1639}{15}e^{-8}=0.9633451168, \qquad
\frac{I^{(2s)}_2}{I^{(FE)}_2}=1-\frac{871}{15}e^{-8}=0.9805208034,
$$
$$
\frac{I^{(2p)}_1}{I^{(FE)}_1}=1-\frac{2101}{45}e^{-8}=0.9843376226, \qquad
\frac{I^{(2p)}_2}{I^{(FE)}_2}=1-\frac{103}{15}e^{-8}=0.9976964900.
$$
The above results indicate a clear tendency: the larger $n$ and $l$, the
closer $I_1^{(nl)}$ and $I_2^{(nl)}$ are to the free-electron values. The
departure from the free-electron behavior does not exceed several percent
at most. These observations provide a solid base for the semiclassical
approach described below.

\subsection{Semiclassical approach}
\label{WKB}
In the one-electron approximation, the Hamiltonian has the form
$H=p^{2}/2m_e + V(r)$, and the density-density Green's function from
Eq.(\ref{fdef}) can be written as
\begin{eqnarray}
\label{f1} F(T,{q}^{2})&=& \left \langle 0 \left |\, e^{-i {\bf q} \cdot
{\bf r}} \, \left [ T -H({\bf p}, {\bf r}) + E_0 \right ]^{-1} \, e^{i
{\bf q}
\cdot {\bf r}}\, \right | 0 \right \rangle \nonumber \\
&=&\left \langle 0 \left | \, \left [ T -H({\bf p} + {\bf q}, {\bf
r}) + E_0  \right ]^{-1} \,  \right | 0 \right \rangle \nonumber\\
&=&
 \left \langle 0 \left | \, \left [ T -{{q}^{2} \over 2 m_e}- {{\bf p}
 \cdot {\bf q} \over m_e}
 - H({\bf p}, {\bf r}) + E_0  \right ]^{-1} \,  \right | 0 \right \rangle
\nonumber \end{eqnarray}
where the infinitesimal shift $T \to T - i \epsilon$ is implied.

Clearly, a nontrivial behavior of the latter expression in Eq.(\ref{f1})
is generated by the presence of the operator $({\bf p} \cdot {\bf q})$ in
the denominator, and the fact that it does not commute with the
Hamiltonian $H$. Thus an analytical calculation of the Green's function as
well as the dynamical structure factor is feasible in only few specific
problems. In Sec.~\ref{H} such a calculation has been presented for
ionization from the $1s$, $2s$, and $2p$ hydrogen-like states. In
particular, we have seen that the deviation of the discussed integrals
(\ref{defi_1}) and (\ref{defi_2}) from their free values are very small:
the largest deviation is exactly at the ionization threshold, where, for
instance, each of the $1s$ integrals is equal to the free-electron value
multiplied by the factor $(1-7 \, e^{-4}/3) \approx 0.957$ (see
Eq.~(\ref{mw1s_num})). It can be also noted from~(\ref{mw1s}) that both
integrals are modified in exactly the same proportion, so that their ratio
is not affected at any $T$: $I_2(T)/I_1(T)= 2 m_e T$. We find however,
that this exact proportionality is specific for the ionization from the
ground state in the Coulomb potential.

 The problem of calculating the integrals
(\ref{defi_1}) and (\ref{defi_2}), however, can be solved in the
semiclassical limit, where one can neglect the noncommutativity of the
momentum ${\bf p}$ with the Hamiltonian, and rather treat this operator as
a number vector. Taking also into account that $(H-E_0) \, |0 \rangle =0$,
one can then readily average the latter expression in Eq.(\ref{f1}) over
the directions of ${\bf q}$ and find the formula for the dynamical
structure factor:
\begin{equation} S(T,{q}^{2})={m_e \over 2 p\,q} \,
\left [ \theta \left ( T- {{q}^{2} \over 2m_e}+{p\,q \over m_e} \right) -
\theta \left ( T- {{q}^{2} \over 2m_e}- {p\,q \over m_e} \right) \right ],
\label{scls}
\end{equation}
where $\theta$ is the standard Heaviside step function. The expression in
Eq.~(\ref{scls}) is nonzero only in the range of $|{\vec q}|$ satisfying
the condition $-p\,q/m_e < T - q^{2}/2m_e < p\, q/m_e$, i.e. between the
(positive) roots of the binomials in the arguments of the step functions:
${q}^{2}_{min}=\sqrt{2m_e T + p^{2}} - p$ and ${q}^{2}_{max}=\sqrt{2m_e T
+ p^{2}} + p$. One can notice that the previously mentioned `spread and
shift' of the peak in the dynamical structure function in this limit
corresponds to a flat pedestal between ${q}^{2}_{min}$ and
${q}^{2}_{max}$. The calculation of the integrals (\ref{defi_1}) and
(\ref{defi_2}) with the expression (\ref{scls}) is straightforward, and
yields the free-electron expressions (\ref{intf_1}) and  (\ref{intf_2})
for the discussed here integrals in the semiclassical (WKB) limit:
\begin{equation} I_1^{(WKB)}={1 \over T}, \qquad I_2^{(WKB)}=2 m_e.
\label{wkbi}
\end{equation}
The appearance of the free-electron expressions here is not surprising,
since the equation (\ref{scls}) can be also viewed as the one for
scattering on an electron boosted to the momentum $p$. The difference from
the pure free-electron case however is in the range of the energy transfer
$T$. Namely, the expressions (\ref{wkbi}) are applicable in this case only
above the ionization threshold, i.e. at $T \ge |E_0|$. Below the threshold
the electron becomes `inactive'.

We believe that the latter conclusion explains the so-called stepping
behavior observed empirically~\cite{kmsf} in the results of numerical
calculations. Namely the calculated cross section $d \sigma/dT$ for
ionization of an electron from an atomic orbital follows the free-electron
dependence on $T$ all the way down to the threshold for the corresponding
orbital with a very small, at most a few percent, deviation. This
observation led the authors of Ref.~\cite{kmsf} to suggest the stepping
approximation for the ratio of the atomic cross section (per target
electron) to the free-electron one:
\begin{equation} f(T) \equiv { d \sigma / dT \over (d \sigma/dT)_{FE}} = {1 \over Z}
\, \sum_i n_i \theta(T - |E_i|), \label{step}
\end{equation}
where the sum runs over the atomic orbitals with the binding energies
$E_i$ and the filling numbers $n_i$. Clearly, the factor $f(T)$ simply
counts the fraction of `active' atomic electrons at the energy $T$, i.e.
those for which the ionization is kinematically possible. For this reason
we refer to $f(T)$ as an {\it atomic factor}. We conclude here that the
stepping approximation is indeed justified with a high accuracy in the
approximation of the scattering on independent electrons, i.e. if one
neglects the two-electron correlations induced by the interference of
terms in the operator $\rho({\bf q})$ in Eq.~(\ref{ne}) corresponding to
different electrons. The effects of such an interference will be discussed
in the next section.

\section{Scattering on manyelectron atoms}
\label{MEA}
In considering the neutrino scattering on actual manyelectron atoms one
needs to evaluate the dependence of the number of active electrons on $T$
and generally also evaluate the effect of the two-electron correlations.
The latter can be studied, for example, in the case of a helium atom,
where the electron-electron correlations are known to play a very
significant role.

\subsection{Helium} \label{He}
Recently, the authors of~\cite{martemyanov11} deduced by means of
numerical calculations that the $\mu_\nu$ contribution to ionization of
the He atomic target by impact of electron antineutrinos from reactor and
tritium sources strongly departures from the stepping approximation,
exhibiting large enhancement relative to the free-electron case. According
to~\cite{martemyanov11}, the effect is maximal when the $T$ value
approaches the ionization threshold in helium, $T_I=24.5874$\,eV, where
the relative enhancement is as large as almost eight orders of magnitude.
It was thus suggested that this finding might have an impact on searches
for $\mu_\nu$, provided that its value falls within the range
$10^{-13}-10^{-12}\mu_B$. In this section we show that (i) the result
of~\cite{martemyanov11} is erroneous and (ii) the stepping approximation
for helium is well applicable, except the energy region $T\sim T_I$ where
the differential cross section substantially decreases relative to the
free-electron case.

We consider the process where an electron antineutrino with energy $E_\nu$
scatters on a He atom at energy and spatial-momentum transfers $T$ and
${\bf q}$, respectively. In what follows we focus on the ionization
channel of this process in the kinematical regime $T\ll E_\nu$, which
mimics a typical situation with reactor ($E_\nu\sim1$\,MeV) and tritium
($E_\nu\sim10$\,keV) antineutrinos when the case $T\to T_I$ is concerned.
The He target is assumed to be in its ground state $|\Phi_i\rangle$ with
the corresponding energy $E_i$. Since for helium one has $\alpha Z\ll1$,
where $Z=2$ is the nuclear charge, the state $|\Phi_i\rangle$ can be
treated nonrelativistically. As we are interested in the energy region
$T\sim T_I$, the final He state $|\Phi_f\rangle$ (with one electron in
continuum) can also be treated in the nonrelativistic approximation.

Under the above assumptions, the dynamical structure factor~(\ref{dsf}) is
given by
\begin{eqnarray}
\label{dyn_str_fact} S(T,q^2)=\sum_f\left|\langle
\Phi_f(\textbf{r}_1,\textbf{r}_2) |e^{i{\bf q}\cdot{\bf r}_1} + e^{i{\bf
q}\cdot{\bf r}_2}|\Phi_i(\textbf{r}_1,\textbf{r}_2)\rangle\right|^2
\delta(T-E_f+E_i).
\end{eqnarray}
Here the $f$ sum runs over all final He states having one electron ejected
in continuum, with $E_f$ being their energies.

For evaluation of the dynamical structure factor~(\ref{dyn_str_fact}) we
employ the same models of the initial and final He states as
in~\cite{martemyanov11}. The initial state is given by a product of two
$1s$ hydrogenlike wave functions with an effective charge $Z_{i}$,
\begin{eqnarray}
\label{init_state}\Phi_i(\textbf{r}_1,\textbf{r}_2)=\varphi_{1s}(Z_{i},\textbf{r}_1)\varphi_{1s}(Z_{i},\textbf{r}_2),
\qquad \varphi_{1s}(Z_{i},\textbf{r})=\sqrt{\frac{Z_{i}^3}{\pi
a_0^3}}\,e^{-Z_{i}r/{a_0}},
\end{eqnarray}
where $a_0=1/(\alpha m_e)$ is the Bohr radius. The final state has the
form
\begin{equation}
\label{fin_state} \Phi_f(\mathbf{r}_1,\mathbf{r}_2)=\frac{1}{\sqrt{2}}[
\varphi_{\mathbf{k}}^-(Z_f,\mathbf{r}_1)\varphi_{1s}(Z,\mathbf{r}_2)+\varphi_{\mathbf{k}}^-(Z_f,\mathbf{r}_2)\varphi_{1s}(Z,\mathbf{r}_1)],
\end{equation}
where $\varphi_{\mathbf{k}}^-(Z_f,\mathbf{r})$ is an outgoing Coulomb wave
for the ejected electron with spatial momentum ${\bf k}$. $Z_f$ is the
effective charge experienced by the ejected electron in the field of the
final He$^+$ ion. Contributions to the dynamical structure factor from
excited He$^+$ states are neglected due to their very small overlap with
the $K$-electron state in the He atom.

To avoid nonphysical effects connected with nonorthogonality of
states~(\ref{init_state}) and~(\ref{fin_state}), we use the Gram-Schmidt
orthogonalization
$$
|\Phi_f\rangle\rightarrow|\Phi_f\rangle-\langle\Phi_i|\Phi_f\rangle|\Phi_i\rangle.
$$
Substitution of~(\ref{init_state}) and~(\ref{fin_state})
into~(\ref{dyn_str_fact}) thus yields
\begin{eqnarray}
\label{dyn_str_fact_1}S(T,q^2)=\int\frac{d{\bf k}}{(2\pi)^3} |F({\bf
k},{\bf q})|^2
\delta\left(T-\frac{k^2}{2m_e}+2\alpha^2m_e-Z_{i}^2\alpha^2m_e\right),
\end{eqnarray}
where $k=\sqrt{2m_e(T+2\alpha^2m_e-Z_{i}^2\alpha^2m_e)}$, and
\begin{equation}
F({\bf k},{\bf q})=\sqrt{2}\langle \varphi_{\bf
k}^-(Z_f,\textbf{r}_1)\varphi_{1s}(Z,\textbf{r}_2)| e^{i{\bf q}\cdot{\bf
r}_1} + e^{i{\bf q}\cdot{\bf r}_2}-2\rho_{1s}({\bf
q})|\varphi_{1s}(Z_{i},{\bf r}_1)\varphi_{1s}(Z_{i},{\bf r}_2)\rangle
\end{equation}
is the inelastic form factor, with
\begin{equation}
\rho_{1s}({\bf q})=\int\varphi_{1s}(Z_{i},{\bf r})e^{i{\bf q}\cdot{\bf
r}}\varphi_{1s}(Z_{i},{\bf r})\,d{\bf r}.
\end{equation}
It is straightforward to perform the further calculation of the dynamical
structure factor analytically (see, for instance, the textbook~\cite{ll}).
The resulting expression is
\begin{equation}
\label{struc_factor_1}S(T,q^2)=\frac{2^{16}\alpha^4m_e^5Z_fZ_{i}^6}{(1-e^{2\pi\eta})(2+Z_{i})^6}
\left[A_1(k,q)+B(k,q)A_2(k,q)+B^2(k,q)\right],
\end{equation}
where $\eta=-\alpha Z_fm_e/k$ and (introducing $p_{i}=\alpha Z_{i}m_e$)
\begin{eqnarray}
A_1(k,q)&=&\frac{\exp\left(2\eta\,{\rm
arccos}\frac{p_{i}^2+q^2-k^2}{\sqrt{(p_{i}^2+q^2+k^2)^2-4k^2q^2}}\right)}{[(p_{i}^2+q^2+k^2)^2-4k^2q^2]^3}
\Bigg\{(p_{i}+\eta k)^2(p_{i}^2+q^2+k^2)^2\nonumber
\\&{}&+4kq^2\left[\frac{1}{3}kp_{i}^2-\frac{2}{3}\eta^2kp_{i}^2-\eta^2k^3-\eta p_{i}(p_{i}^2+q^2+k^2)\right]
\Bigg\},
\end{eqnarray}
\begin{eqnarray}
A_2(k,q)&=&\frac{2\exp\left(\eta\,{\rm
arccos}\frac{p_{i}^2+q^2-k^2}{\sqrt{(p_{i}^2+q^2+k^2)^2-4k^2q^2}}\right)}{(p_{i}^2+q^2+k^2)^2-4k^2q^2}
\left[p_{i}\cos\left(\frac{\eta}{2}\ln\frac{(k+q)^2+p_{i}^2}{(k-q)^2+p_{i}^2}\right)\right.\nonumber\\&{}&
\left.+\frac{p_{i}^2-q^2+k^2}{2q}
\sin\left(\frac{\eta}{2}\ln\frac{(k+q)^2+p_{eff}^2}{(k-q)^2+p_{i}^2}\right)\right],
\end{eqnarray}
\begin{eqnarray}
B(k,q)=e^{2\eta\,{\rm arctg}\frac{k}{p_{i}}}\frac{(Z_{i}-Z_f)\alpha
m_e}{(k^2+p_{i}^2)^2}\left\{\frac{(2+Z_{i})^4\alpha^4m_e^4}{[(2+Z_{i})^2\alpha^2m_e^2+q^2]^2}-\frac{32p_{i}^4}{(4p_{i}^2+q^2)^2}\right\}.\nonumber\\
\end{eqnarray}

Finally, the usual choice of the effective charges is
$Z_i=27/16\approx1.69$ and $Z_f=1$ (see, for instance,~\cite{pra12} and
references therein). The value $Z_i=27/16$ follows from the variational
procedure that minimizes the ground-state energy $E_i$, while the value
$Z_f=1$ ensures the correct asymptotic behavior of the final state.
However, the authors of~\cite{martemyanov11} utilized in their
calculations the values $Z_i=1.79$ and $Z_f=1.1$ derived from fitting the
photoionization cross-section data on helium with the present model of the
He states.

The departures of the differential cross sections~(\ref{d1s})
and~(\ref{d1sw}) from the free-electron approximation are characterized by
the respective atomic factors
\begin{eqnarray}
\label{at_factor} f_{\rm SM}=\frac{d\sigma_{\rm SM}/dT}{d\sigma_{\rm
SM}^{\rm FE}/dT}, \qquad f_{\rm
NMM}=\frac{d\sigma_{(\mu)}/dT}{d\sigma_{(\mu)}^{\rm FE}/dT},
\end{eqnarray}
where $d\sigma_{\rm SM}^{\rm FE}/dT$ and $d\sigma_{(\mu)}^{\rm FE}/dT$ are
the SM and $\mu_\nu$ contributions to the differential cross section for
scattering of an electron antineutrino on two free electrons. Let us
recall that following~\cite{martemyanov11} one should expect the $f_{\rm
NMM}$ value to be of about $10^8$ at $T\to T_I$.
\begin{figure}
\begin{center}
\includegraphics[width=0.9\textwidth]{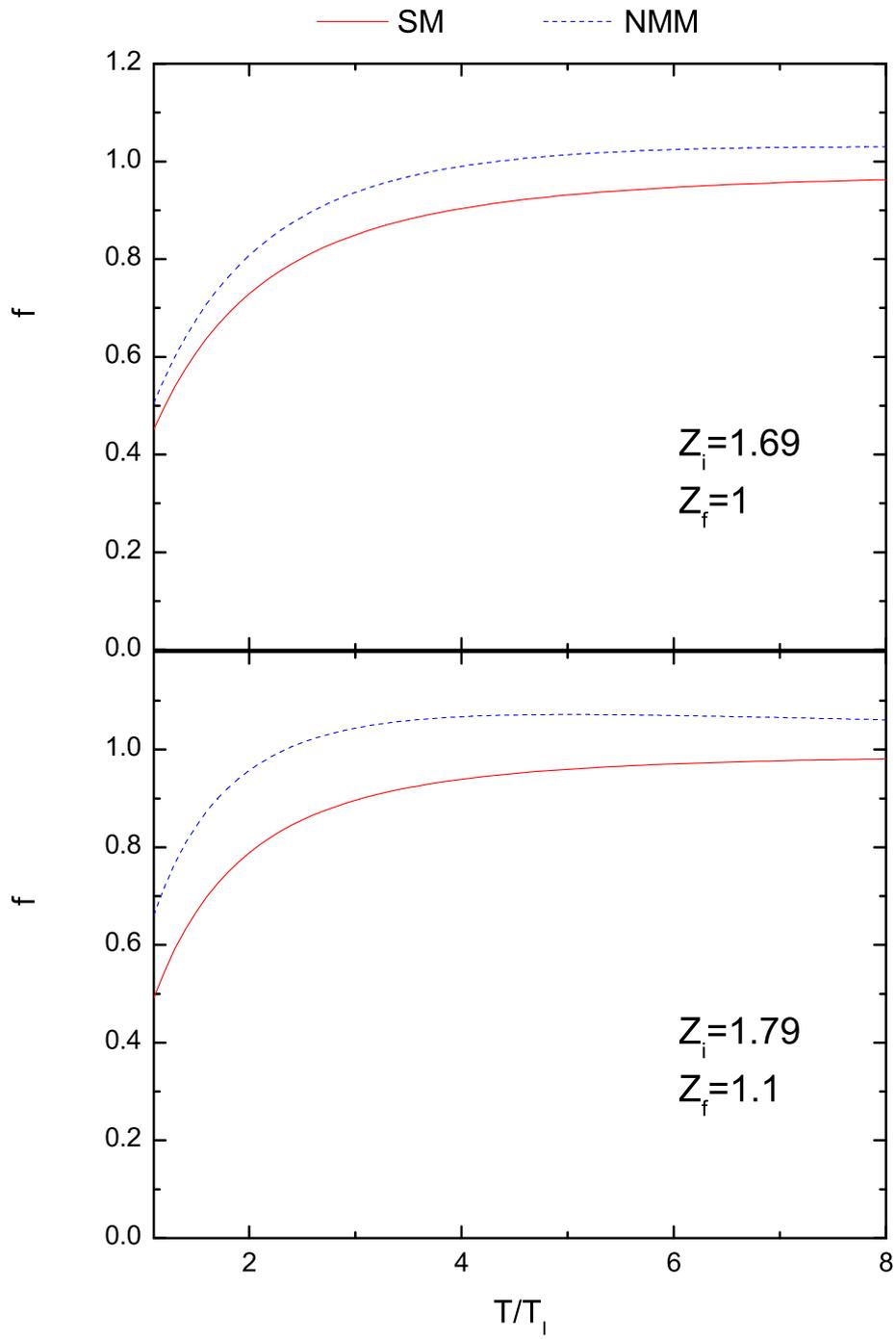}
\end{center}
\caption{\label{fig} Atomic factors~(\ref{at_factor}) as functions of the
energy transfer~\cite{pepan_lett14}.}
\end{figure}

Numerical results for atomic factors~(\ref{at_factor}) are shown in
Fig.~\ref{fig}. They correspond to the kinematical regime $T\ll\alpha
m_e\ll 2E_\nu$, which is typically realized both for reactor and for
tritium antineutrinos when $T<200$\,eV. Note that in such a case one can
safely set the upper limit of integrals in~(\ref{d1s}) and~(\ref{d1sw}) to
infinity, as the dynamical structure factor $S(T,q^2)$ rapidly falls down
when $q\gtrsim\alpha m_e$ and practically vanishes in the region
$q\gg\alpha m_e$. It can be seen from Fig.~\ref{fig} that atomic factors
exhibit similar behaviors for both sets of the $Z_i$ and $Z_f$ parameters
discussed in the previous section. Namely, their values are minimal
($\sim0.5$) at the ionization threshold, $T=T_I$, and tend to unity with
increasing $T$. The latter tendency is readily explained by approaching
the free-electron limit. It can be also seen that a more or less serious
deviation ($>10\%$) of the present results from the stepping approximation
is observed only in the low-energy region $T<100$\,eV. This deviation can
be attributed to the effect of the electron-electron correlation in a
helium atom. Indeed, if the electrons do not interact with each other,
then they occupy two $1s$ hydrogen-like states (with opposite spins), in
which case the departure of the atomic-factor values from unity is,
according to the results of Sec.~\ref{H}, less than 5\%.

Thus, the calculations presented in Fig.~\ref{fig} do not confirm the huge
enhancement of the $\mu_\nu$ contribution with respect to the
free-electron approximation. Moreover, in accord with various calculations
for other atomic targets~\cite{ddf92,fdd92,kmsf,fms,kms03,ks,jpl11,prd11},
we find that at small energy-transfer values the electron binding in
helium leads to the appreciable reduction of the differential cross
section relative to the free-electron case. We attribute the erroneous
prediction of~\cite{martemyanov11} to the incorrect dynamical model that
draws an analogy between the NMM-induced ionization and photoionization.
Indeed, as discussed in Sec.~\ref{nu-at_el}, the virtual photon in the
NMM-induced ionization process can be treated as real only when $q\to T$.
However, the integration in~(\ref{d1s}) involves the $q$ values ranging
from $T$ up to $2E_\nu-T$. Since $E_\nu\gg T$, the real-photon picture
appears to be applicable only in the vicinity of the lower integration
limit. When moving away from that momentum region, one encounters a strong
departure from the real-photon approximation which treats the integrand as
a constant in the whole integration range, assuming it to be equal to its
value at $q=T$, that is,
$$
\frac{1}{q^2}\,S(T,q^2)=\frac{1}{T^2}\,S(T,T^2).
$$
Such an approach is manifestly unjustified, and it gives rise to the
spurious enhancement of the $\mu_\nu$ contribution to the differential
cross section.
\subsection{Thomas-Fermi model} \label{TF}
\begin{figure}
\begin{center}
    \includegraphics[width=0.9\textwidth]{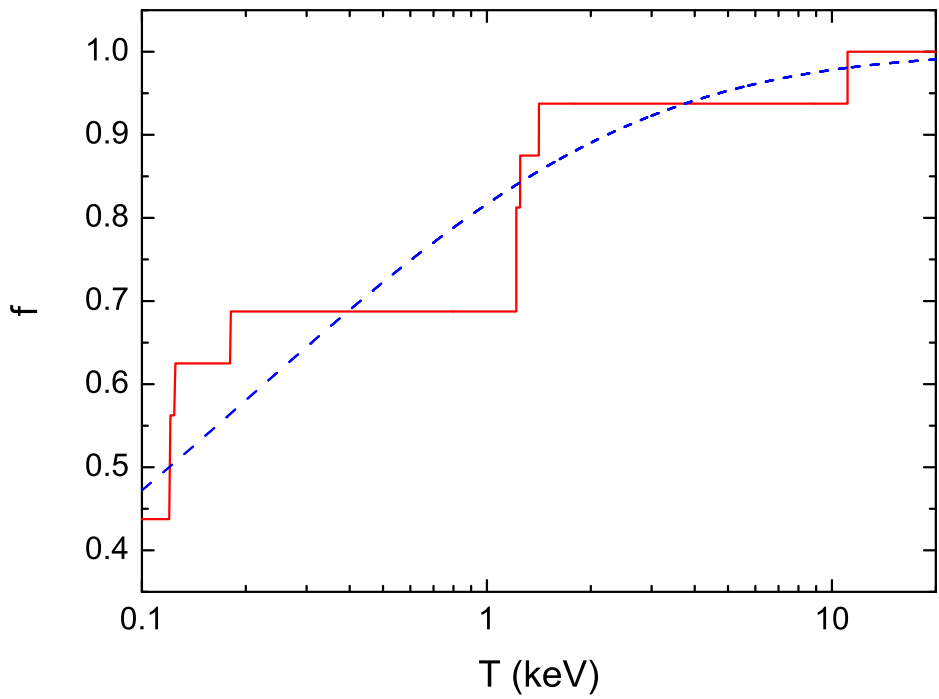}
   \caption{\label{fig2} The atomic factor $f$ for germanium in the stepping approximation with the actual energies of the
   orbitals (solid line) and its interpolation in the Thomas-Fermi model
   (dashed)~\cite{jpl11}.}
\end{center}
\end{figure}

In the Thomas-Fermi model (see e.g. Ref.~\cite{ll}) the atomic electrons
are described as a degenerate free electron gas in a master potential
$\phi(r)$ filling the momentum space up to the zero Fermi energy, i.e. up
to the momentum $p_0(r)$ such that $p_0^2/2 m_e - e \phi=0$. The electron
density $n(r)=p_0^3/(3 \pi^2)$ then determines the potential $\phi(r)$
from the usual Poisson's equation. In the discussed picture at an energy
transfer $T$ the ionization is possible only for the electrons whose
energies in the potential are above $-T$, i.e. with momenta above $p_T(r)$
with $p_T^2/2m_e - e \phi = -T$. The electrons with lower energy are
inactive. Calculating the density of the inactive electrons as $p_T^3/(3
\pi^2)$ and subtracting their total number from Z, one readily arrives at
the formula for the atomic factor, i.e. the effective fraction of the
active electrons $Z_{\rm eff}/Z$ as a function of $T$:
\begin{equation} f(T)={Z_{\rm eff}(T) \over Z}= 1 - \int_0^{x_0(T)} \, \left [
{\chi(x) \over x} - {T \over T_0} \right ]^{3/2} \, x^2 \, dx~,
\label{zeff}
\end{equation}
where $\chi(x)$ is the Thomas-Fermi function, well known and tabulated, of
the scaling variable $x = 2 (4/3\pi)^{2/3} m_e \alpha Z^{1/3}$,  the
energy scale $T_0$ is given by
\begin{equation}
T_0=2 \, \left ( {4 \over 3 \pi} \right )^{2/3} \, m_e \, \alpha^2 \,
Z^{4/3} \approx  30.8 \, Z^{4/3}\, {\rm eV}~, \label{t0}
\end{equation}
and, finally, $x_0(T)$ is the point where the integrand becomes zero, i.e.
corresponding to the radius beyond which all the electrons are active at
the given energy $T$. The energy scale $T_0$ in germanium (Z=32) evaluates
to  $T_0 \approx 3.1$\,keV. The Thomas-Fermi atomic factor for germanium
calculated from the formula (\ref{zeff}) is shown by the dashed line in
the plot of Fig.~\ref{fig2}. The discussed statistical model is known to
approximate the average bulk properties of the atomic electrons with a
relative accuracy $O(Z^{-2/3})$ and as long as the essential distances $r$
satisfy the condition $Z^{-1} \ll m_e \alpha r \ll 1$, which condition in
terms of the scaling variable $x$ reads as $Z^{-2/3} \ll x \ll Z^{1/3}$.
In terms of the formula (\ref{zeff}) for the number of active electrons,
the lower bound on the applicability of the model is formally broken at $T
\sim Z^{2/3} T_0$, i.e. at the energy scale of the inner atomic shells.
However the effect of the deactivation of the inner electrons is small, of
order $Z^{-1}$ in comparison with the total number $Z$ of the electrons.
On the other hand, at low $T$, including the most interesting region of $T
\sim T_0$, the integral in Eq.~(\ref{zeff}) is determined by the range of
$x$ of order one, where the model treatment is reasonably justified.

The energies of the inner $K,~L$ and $M$ orbitals in the germanium atom
are well known (see e.g. Ref.~\cite{henke93}) and provide the necessary
data for a description of the neutrino scattering by the stepping formula
(\ref{step}) down to the values of the energy transfer $T$ in the range of
the binding of the $M$ electrons, i.e. at $T
> |E_M| \approx 0.18$\,keV. The corresponding steps in the atomic
factor are shown in Fig.~\ref{fig2}. One can see that the stepping atomic
factor~(\ref{step}) mimics upon average over the energy intervals between
the electron shells in germanium the Thomas-Fermi result. Thus, it can be
considered as refinement of the latter due to account for the quantization
of the electron binding energies. It can be mentioned that if one applies
formulas of Sec.~\ref{H} to the onset of the $K$ shell step, i.e. just
above 10.9\,keV, the difference from the shown in the plot step function
would be practically invisible in the scale of Fig.~\ref{fig2}.

\subsection{\emph{Ab initio} approaches}
While the treatments based on a generic model of manyelectron atomic
targets allow to determine characteristic features and behaviors of the
differential cross sections~(\ref{d1s}) and~(\ref{d1sw}), to obtain
accurate numerical results one needs to resort to \emph{ab initio}
calculations. Such calculations can be realized using the Hartree-Fock
(HF) method (see, for instance,~\cite{ll}) and its modifications. In the
HF approximation, atomic electrons occupy one-electron states in a
spherically-symmetric mean-field potential which is derived
self-consistently from the solution of the HF equations. Accordingly, each
one-electron state independently contributes to the atomic ionization
process. For the first time this approach was formulated in
Refs.~\cite{ddf92,fdd92}, where it was illustrated with numerical
calculations of neutrino-impact ionization of the F and Mo atomic targets.
The wave functions and energies of atomic bound states were calculated
within the relativistic HF method~\cite{grant61,grant65} with local
exchange-correlation potential~\cite{moruzzi_book}. The wave functions of
outgoing electrons were obtained by a numerical solution of the Dirac
equation in the same mean-field potential as for the wave functions of
discrete states. Performed in Ref.~\cite{kmsf} numerical calculations for
ionization of the iodine atoms by impact of reactor antineutrinos led the
authors to suggest the stepping approximation~(\ref{step}).

In a very recent theoretical study~\cite{chen14}, the authors adopted the
multiconfiguration relativistic random-phase approximation
(MCRRPA)~\cite{huang82_1,huang82_2} to evaluate the germanium atomic
factors. This particular method is based on the time-dependent HF
approximation~\cite{jorgensen75}, however, several important features make
it a better tool beyond the usual HF approximation to describe transitions
of open-shell atoms of high atomic number $Z$. First, for open-shell
atoms, typically there are more than one configuration which have the
desired ground-state properties. Therefore, a proper HF reference state
should be formed by a linear combination of these allowed configurations,
i.e., a multiconfiguration reference state. Second, for atoms of high $Z$,
the relativistic corrections can no longer be ignored. By using a Dirac
equation, instead of a Schr\"odinger one, the leading relativistic terms
in the atomic Hamiltonian are treated nonperturbatively from the onset.
Third, two-body correlations in addition to the HF approximation are
generally important for excited states and transition matrix elements. The
random-phase approximation (RPA) is devised to account for part of the
additional two-body correlations (particles can be in the valence or core
states) not only for the excited but also for the reference state, and in
a lot of cases, it gives good agreement with experiment~\cite{amusia75}.
Furthermore, it has been shown that RPA equations preserve gauge
invariance~\cite{lin77}; this provides a measure of stability of their
solutions.

The MCRRPA has been applied successfully to photoexcitation and
photoionization of divalent atoms such as Be, Mg, Zn, and etc. (some of
the results are summarized in~\cite{huang95}). Following similar
treatments, the authors of~\cite{chen14} treated the electronic
configuration of germanium as a core filled up to the 4$s$ orbits, with
two valence electrons in the $4p$ orbits. As the Ge ground state is a
$^3P_0$ state, it is a linear combination of two configurations, namely
[Zn]$4p^2_{1/2}$ and [Zn]$4p^2_{3/2}$. The wave function was calculated
using the multiconfiguration Dirac-Fock package~\cite{desclaux75}. The
atomic excitations due to weak and magnetic scattering were solved by the
MCRRPA equation, and consequently transition matrix elements were yielded.
In that calculation, all the current operators were expanded by spherical
multipoles, and the resulting final scattering states were represented in
the spherical wave basis and subject to the incoming-wave boundary
condition.

Compared with the previous works on the same
subject~\cite{ddf92,fdd92,kmsf,fms,kms03} which are also in the similar
spirit of the relativistic HF method, the MCRRPA approach differs in
several respects. First, due to the near degeneracy of the
$N_{II}(4p_{3/2})$ and $N_{III}(4p_{1/2})$ levels, using a
multiconfiguration reference state is necessary. Second, the nonlocal Fock
term is treated exactly, without resorting to the local exchange
potentials. Third, the excited states are calculated with two-body
correlation built in by MCRRPA, not simply by solving a Coulomb wave
function with a static one-hole mean field.

\begin{figure}
\begin{center}
    \includegraphics[width=0.9\textwidth]{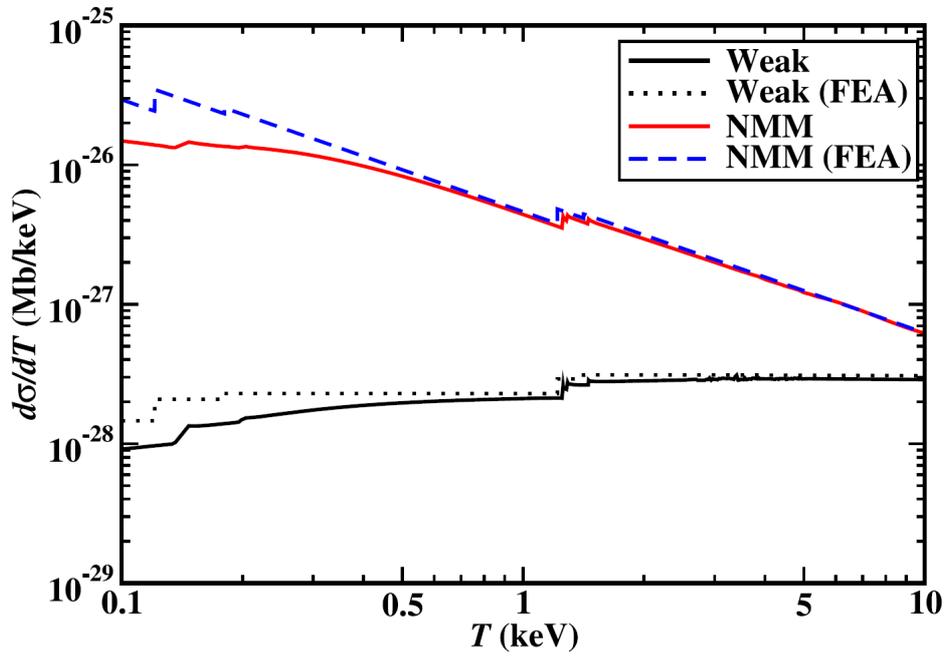}
   \caption{\label{fig:Ge_ab_initio} The SM (weak) and NMM contributions to the differential cross section of $\bar{\nu}_e$-Ge ionization at
   $E_\nu=1$\,MeV~\cite{chen14} in comparison with the corresponding stepping-approximation results (FEA).
   The NMM value is set to be the current upper limit $\mu_\nu=2.9\times10^{-11}\mu_B$~\cite{GEMMA}.}
\end{center}
\end{figure}
Fig.~\ref{fig:Ge_ab_initio} shows numerical results from
Ref.~\cite{chen14} for ionization of germanium by impact of an electron
antineutrino. As can be seen, in the energy region $T\gtrsim1$\,keV the
results are very well explained by the stepping-approximation
formula~(\ref{step}). At the same time, in the sub-keV region, i.e., where
the electrons from the $K$ and $L$ shells in germanium stay `inactive',
both the SM and NMM contributions appear to be significantly suppressed as
compared to the stepping approximation. The latter finding seems to
disagree with the semiclassical approach discussed in Sec.~\ref{WKB},
according to which the ionization involving more loosely-bound electron
states, such as those belonging to the $M$ and $N$ shells, is expected to
follow more closely the free-electron scenario. Notice that a similar
suppression of the atomic-factor values close to the ionization threshold
was observed in the case of helium (see Fig.~\ref{fig}), and it was
attributed to the two-electron correlation effect. Thus, we can suggest
that the correlation effects beyond the approximation of independent
electrons lead to the suppression of atomic factors in the low-energy
region. This feature will be important for the next-generation experiments
with Ge detectors having energy thresholds in the sub-keV
region~\cite{wong11,li13,li14}.

%
\section{Summary and perspectives}
\label{summary}
In this review, we have considered the neutrino-atom ionizing collisions
with focus on the most important theoretical issues related to the
problem. The main results discussed in the paper can be summed up as
follows.

The differential over the energy transfer cross section given by the
free-electron formulas (\ref{cr_sec_SM}) and (\ref{cr_sec_mu}) and the
stepping behavior of the atomic factor given by Eq.~(\ref{step}) provides
a reasonable description of the neutrino-impact ionization of a complex
atom, such as germanium, down to quite low energy transfer. The deviation
from this approximation due to the onset of the ionization near the
threshold is less than 5\% (of the height of the step) for the $K$
electrons, if one applies the analytical behavior of this onset that one
finds for the ground state of a hydrogen-like ion. It is also found that
the free-electron expressions for the differential cross section are not
affected by the atomic binding effects in the semiclassical limit and for
independent electrons. These analytical results can support the
numerically determined behaviors of the electroweak and magnetic
contributions to the neutrino-impact ionization of various atomic targets
within the mean-field model~\cite{ddf92,fdd92,kmsf,fms}. At the same time,
very recent numerical calculations of the $\bar{\nu}_e$-impact ionization
processes of helium~\cite{pepan_lett14} and germanium~\cite{chen14}
exhibit suppression of the SM and NMM differential cross sections relative
to the stepping approximation with lowering the energy-transfer value.
This suppression can be assigned to the electron-electron correlation
effects.

A theoretical analysis~\cite{pepan_lett14} of ionization of helium by
electron-antineutrino impact shows no evidence of the recently predicted
enhancement~\cite{martemyanov11} of the electromagnetic contribution as
compared with the free-electron case. In contrast, in line with previous
studies on other atomic targets, it is found that the magnitudes of the
differential cross sections decrease relative to the free-electron
approximation when the energy transfer is close to the ionization
threshold. Thus, no sensitivity enhancement can be expected when using the
He atomic target in searches for NMM. And the stepping approximation
appears to be valid, within a few-percent accuracy, down to the
energy-transfer values as low as almost 100\,eV. We thus conclude that for
practical applications, i.e. for the analysis of data of the searches for
NMM, one can safely apply the free-electron formulas and the stepping
approximation at the energy transfer down to this range.

When analyzing the low-$T$ data of the current high-sensitivity
experiments searching for neutrino electromagnetic properties, one must go
beyond the free-electron approximation for the elastic neutrino-electron
scattering and take into account the atomic-ionization effects, at least,
in the case of $K$ electrons. At the present time, the experiment GEMMA-II
with reactor antineutrinos is in preparation~\cite{GEMMA}. Its sensitivity
to the NMM value is expected to be at the level of
$1\times10^{-11}\,\mu_B$. To achieve such a sensitivity level, which is
the region of astrophysical interest~\cite{raffelt90}, it is planed to
reduce the effective energy threshold of a Ge detector from 2.8 to
1.5\,keV. This threshold value will be very close to the binding energies
of the $L$ electrons in germanium (1.2-1.4\,keV~\cite{henke93}). Recently,
a $p$-type point-contact Ge detector~\cite{wong11,li13,li14} has been
implemented in the TEXONO experiment with reactor antineutrinos. The
energy threshold of this detector is about 0.3\,keV, which value is
comparable to the binding energies of the $M_{I-III}$ electrons in
germanium (0.12-0.18\,keV~\cite{henke93}). This means that an accurate
analysis of the corresponding data will require numerical calculations
based on the \emph{ab initio} methods.

With lowering the $T$ value down to $T=2E_\nu^2/(E_\nu+M_N)$, an
additional collision channel apart from ionization opens up, namely the
coherent elastic neutrino-nucleus scattering~\cite{freedman74}, which has
not been observed experimentally so far. The early treatments of the
atomic effects in the coherent elastic neutrino-nucleus scattering within
SM can be found in~\cite{gdt75,gt77,sw86}. It should be noted that any
deviation of the measured cross section of the coherent elastic
neutrino-nucleus scattering from the very precisely known SM
value~\cite{ds84} will provide a signature of the BSM physics
(see~\cite{scholberg06,barranco05,barranco07,davidson03}). In this
context, the accurate calculations of the NMM-induced contribution to the
cross section of the coherent elastic neutrino-nucleus scattering appear
to be of particular importance. The NMM-induced coherent neutrino
scattering by single atoms as well as by crystals was discussed
theoretically only in~\cite{amg90}. However, further studies are necessary
for the correct interpretation of future measurements at low $T$
values~\cite{ffa12}.

\section*{Acknowledgements}
This work was partially supported by the Russian Foundation for Basic
Research (Grants no. 14-01-00420-a and no. 14-22-03043-ofi-m).

\end{document}